\documentclass[10pt]{extarticle}
\usepackage[utf8]{inputenc}
\usepackage[normalem]{ulem}% To strikethrough some text
\usepackage[]{amsmath} %felqn
\usepackage{amssymb}
\usepackage{nccmath} % temporarilly centered equation
\usepackage{gensymb} % needed for \degree
\usepackage{bm}
\usepackage{colortbl}  % cellcolor
\usepackage[table,dvipsnames]{xcolor}    % loads also »colortbl«
\usepackage{graphicx, wrapfig, subcaption, setspace, booktabs}
\usepackage{subcaption} % For subfigures
\usepackage{float}
\usepackage{parskip}
\usepackage{empheq}
\usepackage[theorems,skins]{tcolorbox}
\usepackage{xcolor}
\usepackage{tikz}
\usepackage{natbib}
\usepackage{wrapfig}   % Für die wrapfigure-Umgebung
\usepackage{comment}
\usepackage[a4paper, total={14cm, 24cm},]{geometry}
\usepackage{layout}    %  um die Seitenränder als Bild auszugeben
\usepackage{empheq}
\usepackage{xspace} % xspace used in glossaries
\usepackage{myglossaries}
\usepackage[colorlinks = true,
            linkcolor = blue,
            urlcolor  = blue,
            citecolor = blue,
            anchorcolor = blue]{hyperref}
\begin{document}

% =============================================================================
% =============================================================================
%   TITLE
% =============================================================================
% =============================================================================
\title{\LARGE\textbf{Hydrodynamic interactions of  self-propelled \\ 
tandem flappers with passive pitching}}
\author{Maximilian Schenk$^{a}$, Gonzalo Arranz$^{b}$, Oscar Flores$^{c}$, \\
Manuel Moriche$^{a}$ and Manuel Garc\'ia-Villalba$^{a}$}
\date{\vspace{-2ex}}
\maketitle
\begin{center}
\small
$^{a}$\textit{Institute of Fluid Mechanics and Heat Transfer, 
TU Wien, Vienna, Austria} \\
$^{b}$\textit{Graduate Aerospace Laboratories, 
Caltech, Pasadena, California, USA} \\
$^{c}$\textit{Aerospace Engineering Department, 
Universidad Carlos III de Madrid, Spain}\\
\end{center}

% =============================================================================
% =============================================================================
%   ABSTRACT
% =============================================================================
% =============================================================================
\rule{\textwidth}{0.6pt}
\vspace{-0.2cm}

\begin{abstract}
We simulate two in-line, self-propelled three-dimensional flappers of finite 
span, driven by sinusoidal heaving motions with a phase difference \gls{dphi}. 
Pitching is passive, governed by fluid--structure interaction and resisted by 
a torsional spring. 
We first consider identical spring stiffnesses, set to the value that maximises 
the propulsive speed of an isolated flapper.
Allowing the pair to self-arrange, we characterise the equilibrium distance, 
propulsive speed, input power, and efficiency over the full range of phase 
shifts.
The pair settles into a stable equilibrium for all phase shifts, travelling at 
a common propulsive speed and maintaining an equilibrium distance that 
increases approximately linearly with \gls{dphi}.
The leader is largely unaffected by the presence of the follower and performs 
similarly to an isolated flapper, whereas the follower requires more power in 
every case.
This additional power outweighs the small increase in propulsive speed, so that 
the tandem generally provides no efficiency benefit over an isolated flapper, 
with comparable efficiency reached only at an intermediate phase shift.
For equal stiffness, the power penalty of the follower is associated with a 
reduction of its pitching amplitude by the wake of the leader, while its 
pitching phase remains nearly unchanged relative to the isolated flapper.
Keeping the leader at the stiffness that maximises its isolated propulsive 
speed, we investigate whether the tandem performance can be improved by 
independently tuning the follower through the stiffness ratio \gls{knr} 
in the range $0.8 \le\gls{knr} \le 1.2$.
Reducing the follower stiffness lowers its power requirement and increases 
the tandem efficiency by an almost phase-independent amount, without affecting 
the equilibrium distance or propulsive speed significantly.
For the softest follower considered, the tandem is more efficient than an 
isolated flapper across the full range of phase shifts, with gains of about 
2--3\% for $\gls{dphi} \ge 90^\circ$.
\end{abstract}
\rule{\textwidth}{0.6pt}

% =============================================================================
% =============================================================================
%   INTRODUCTION
% =============================================================================
% =============================================================================
\section{Introduction}
\label{sec:intro}

% -----------------------------------------------------------------------------
Collective locomotion in nature, as observed in schooling fish and flocking
birds, provides a compelling example of how hydrodynamic interactions between
bodies can significantly influence their propulsive performance
\citep{ligman2023,ko2023,timm2024,flores2025, kanso2026}.
In such configurations, each body operates within a flow field modified by its
neighbours, resulting in collective effects that may be exploited to enhance
propulsion compared with that of an isolated body.
% -----------------------------------------------------------------------------
\citet{weihs1973} predicted that fish arranged in a diamond pattern could
reduce the cost of swimming by exploiting the velocity induced by the vortex
wakes of the individuals ahead of them.
While the biological drivers of schooling behaviour remain an active area of
research, the hydrodynamic benefits of collective locomotion have been
demonstrated in a range of biological and robotic systems
\citep{weimerskirch2001, becker2015}.
A moving swimmer sheds a train of vortices, whose induced velocity at a
trailing body may align with its motion or oppose it.
Depending on this alignment, the same vortices can reduce or increase the
power required by a follower, so the outcome depends on the timing of the
interaction \citep{boschitsch2014, li2020}.
A trout swimming behind a cylinder, for instance, adapts its body motion to
the oncoming vortices to reduce its muscular activity, a behaviour known as
the K\'arm\'an gait \citep{liao2003fish,liao2003}.
\citet{beal2006} observed that a dead fish held behind a cylinder was able
to overcome its own drag when its flexible body resonated with the oncoming
vortices.
This result showed that favourable interactions with an incoming wake depend
not only on the wake itself, but also on the passive dynamic response of the
body to the unsteady forcing.
When the bodies are fixed in position in a free stream, the conditions of
their interaction can be controlled.
The separation and the phase between the prescribed motions are independent
parameters that can be tuned so that the follower extracts energy from the
oncoming vortices \citep{muscutt2017, kurt2018}.
For tethered three-dimensional fish-like swimmers, \citet{zhou2025} found
that changing the phase difference mainly shifts the regions of thrust
enhancement in the streamwise direction, so that separation and phase act
as interchangeable parameters.
% -----------------------------------------------------------------------------

% -----------------------------------------------------------------------------
For two flapping foils in tandem, most studies have considered configurations
in which both foils undergo prescribed motion in a uniform free stream.
This setup allows the relative motion of both bodies to be controlled
independently, making it possible to investigate the effects of parameters
such as the inter-foil distance and the phase shift between their motions.
\citet{boschitsch2014} showed that the performance of the leading foil is barely
affected by the presence of the follower, whereas the thrust and efficiency of
the follower vary strongly with the streamwise gap and the phase difference.
Depending on these two parameters, the performance of the follower can improve
or worsen relative to that of an isolated foil.
Similar trends have been reported for other tandem configurations
\citep{muscutt2017, kurt2018}.
% -----------------------------------------------------------------------------

% -----------------------------------------------------------------------------
When two foils in tandem are free to self-propel, their streamwise motion
results from the balance of hydrodynamic forces.
The separation between the leader and the follower emerges from their
hydrodynamic interaction and the flow encountered by the follower cannot
be controlled independently.
This self-organisation has been observed in a range of two-dimensional
configurations involving self-propelled foils, both in pure heaving motion and
in combined heaving and pitching motion
\citep{zhu2014, peng2018, lin2019, ryu2020}.
In experiments with freely swimming flapping wings, \citet{becker2015} and
\citet{ramananarivo2016} showed that the pair adopts discrete, quantised
equilibrium positions, locked to integer multiples of the wavelength of the
leader's trajectory.
\citet{newbolt2019} found that an imposed phase shift between the motions of
the two bodies displaces this equilibrium distance in proportion to it, so
that the arrangement of the pair is set by the phase at which the follower
meets the leader's wake.
A recent review by \citet{kanso2026} highlights that this linear phase--distance
relation is robust across rigid and flexible swimmers, two- and three-dimensional
configurations, and numerical and experimental realisations.
The same review notes that this passive coordination, established for identical
swimmers, is sensitive to mismatches in their flapping kinematics.
Across these studies, the leader's performance remains largely unchanged by the
presence of the follower, whereas the follower is strongly affected by the
leader's wake.
It requires less power when the velocity induced by the leader's vortices aligns
with its motion, and more when it opposes it.
The mechanism by which a follower matches its motion to the oncoming vortices is
known as vortex phase matching \citep{li2020}.
For a side-by-side pair of pitching foils, the gain in efficiency is much
smaller when the pair is self-propelled than when it is held at a fixed
position \citep{ormonde2024}.
All of the numerical self-propelled studies cited above consider
two-dimensional configurations, for which the wake is characterised by a reverse
von K\'arm\'an street of counter-rotating vortices that maintains a narrow and
coherent propulsive jet far downstream.
% -----------------------------------------------------------------------------

% -----------------------------------------------------------------------------
Another aspect of the problem that should be considered is the dimensionality of
the problem.
Because of the high computational cost of three-dimensional simulations, most of
the literature found for flapping tandem configurations assumes a
two-dimensional problem.
Three-dimensional mechanisms such as vortex stretching can appear even when an
infinite wing is considered \citep{moriche2016}.
Furthermore, if one considers wings of finite span, vortical structures such as
wing-tip vortices appear.
Typical observations include a loss of coherence of the wake, resulting in the
momentum of the reverse von K\'arm\'an wake no longer being concentrated on the
centreline, but laterally spread, showing a faster decay
\citep{buchholz2008,gazzola2011}.
As a result, the hydrodynamic benefit is systematically weaker in three
dimensions \citep{arranz2020}.
\citet{kurt2018} compared the same tandem arrangement for two-dimensional foils
and rectangular wings of aspect ratio 2 and found the follower's performance
trends to change once the foils were given finite span.
Fully three-dimensional self-propelled configurations have been examined only
rarely.
In the work of \citet{verma2018}, as in recent simulations of fish-like pairs
\citep{huang2025,zhou2025}, the relative position of the two bodies is
externally controlled rather than left to emerge from the flow.
\citet{ormonde2026} let two three-dimensional tuna-like swimmers move freely in
the streamwise direction and observed a stable side-by-side formation, maintained
by flow channelling between the bodies rather than by vortex--body interactions.
\citet{arranz2022} simulated a freely swimming tandem of two finite-span
chordwise-flexible plates, driven by a prescribed heaving motion at the leading
edge, and found quantised equilibrium positions similar to those in the
two-dimensional studies mentioned before.
% -----------------------------------------------------------------------------

% -----------------------------------------------------------------------------
To our knowledge, all these studies involving self-propelled foils assign
identical structural properties to the leader and the follower.
Yet the stiffness of a passively pitching flapper sets the timing of its pitch
relative to its heave, and thereby the phase at which it meets the vortices
shed by the leader.
There is no reason for the stiffness that is optimal for an isolated swimmer to
remain optimal for a follower immersed in the leader's wake.
For example, in fixed tandem wing systems the downstream body has been found to
perform best when its structural or geometric properties differ from those of
the upstream one, whether in stiffness \citep{martinez2024} or, for thrust,
in aspect ratio \citep{jurado2022}.
The follower can also differ from the leader in its prescribed kinematics:
mismatching their pitching amplitudes raises the collective efficiency both
of a tandem held at a fixed position \citep{kurt2021} and of a self-propelled
tandem, whose formation remains stable \citep{han2025}.
Whether tuning the follower in this way benefits a freely swimming pair, and
whether the resulting formation remains stable, have not been addressed.
% -----------------------------------------------------------------------------

% -----------------------------------------------------------------------------
In the present work we address these questions by means of direct numerical
simulations of a tandem pair of self-propelled, finite-span flappers.
Each flapper is driven by a prescribed heaving motion and is free to rotate
about its centre of gravity, where a torsional spring provides a restoring
torque, so that its pitch arises from the fluid--structure interaction rather
than being imposed.
The torsional stiffness is set to the value that maximises the propulsive
speed of a flapper in isolation \citep{schenk2026}.
With this stiffness fixed, we place a leader and a follower in an in-line
tandem configuration and let the pair self-arrange, characterising the
resulting configurations and their performance over the imposed phase
shifts between the two flappers.
In doing so, we test whether the passive-pitch model captures the propulsive
performance of the chordwise-flexible flappers of \citet{arranz2022}.
We then vary the follower stiffness independently of the leader to determine
whether this improves the performance of the tandem configuration.
% -----------------------------------------------------------------------------

% -----------------------------------------------------------------------------
The remainder of the paper is organised as follows: the problem is presented in
\S~\ref{sec:prob} and the numerical method used to tackle the problem is
described in \S~\ref{sec:meth}.
The results are presented in \S\,\ref{sec:results}, starting from the tandem
with identical torsional spring stiffness and turning afterwards to the effect
of varying the follower's stiffness.
Conclusions are drawn in \S\,\ref{sec:conclusions}.
% -----------------------------------------------------------------------------

% =============================================================================
% =============================================================================
%   METHODOLOGY
% =============================================================================
% =============================================================================
\section{Problem description}
\label{sec:prob}

\begin{figure}
    \centering
    \includegraphics[width=1.0\textwidth]{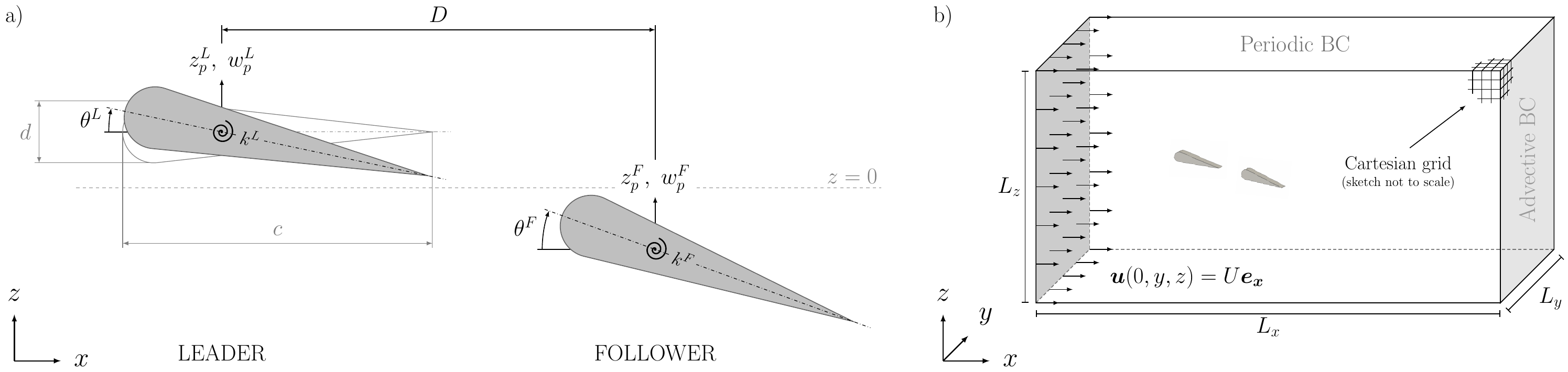}
    \caption{a) Sketch of the problem (to scale) indicating the geometry of the
    flappers, the measure of the distance between them and the actuation point
    of the rotational spring.
    b) Sketch of the computational domain including the boundary conditions 
    (BC).}
    \label{fig:meth:sketch}
\end{figure}

% =============================================================================
%   FLOW
% =============================================================================
We investigate the hydrodynamic interaction of two self-propelled
three-dimensional flappers in a tandem configuration, as shown in
Figure~\ref{fig:meth:sketch}a).
Both flappers are immersed in an otherwise quiescent fluid of constant density
\gls{rhof} and kinematic viscosity \gls{nu}.
The velocity of the flow $\gls{vu}=(\gls{vu_x},\gls{vu_y},\gls{vu_z})$ and the
pressure \gls{p} are governed by the Navier--Stokes equations of the
incompressible flow
\begin{subequations} \label{eq:ns}
\begin{align}
    \nabla  \cdot \gls{vu} &= 0 \, ,  \label{eq:ns:cont}\\
    \ppt{\gls{vu}}
    + \left(\gls{vu} \cdot \nabla \right) \gls{vu}
        &= -\frac{\nabla \gls{p}}{\gls{rhof}} + \gls{nu} \nabla^2 \gls{vu} \, ,
    \label{eq:ns:mom}
        %+ \gls{fibm} \, ,
\end{align}
\end{subequations}
complemented with the no-slip and no-penetration boundary conditions at the
surface of the flappers.
% ------------------------------------------------------------------------------

% =============================================================================
%   PARTICLE
% =============================================================================

% ------------------------------------------------------------------------------
The flappers are geometrically identical rigid bodies with uniform density
\gls{rhop} and volume $\gls{Vp}$.
We denote by \gls{J} the specific moment of inertia (moment of inertia per
unit of mass) about the spanwise axis through the centre of mass.
The cross section of the flappers (uniform along the span, \gls{b}) consists of
a circular leading edge of diameter $\gls{d}$ that blends tangentially into two
straight segments connected to the (sharp) trailing edge.
In this work, the flapper has a maximum thickness equal to the diameter of
its rounded leading edge, $\gls{d}=0.2\gls{c}$, and a span of
$\gls{b}=0.5\gls{c}$, which gives an aspect ratio $\gls{b}/\gls{c}=0.5$.
The resulting cross-sectional area is $A_c=0.106\,\gls{c;sq}$, so that the
volume of the flapper is $\gls{Vp}=A_c\,\gls{b}=0.053\,\gls{c;cb}$.
The centre of mass, \gls{vxp}, is located at a distance of $0.349\,\gls{c}$
from the leading edge and the specific moment of inertia of the flapper is
$\gls{J}=0.055\,\gls{c;sq}$.
Further details of the geometry can be found in
\citet[Appendix A.1]{schenk2026}.
%-------------------------------------------------------------------------------

%-------------------------------------------------------------------------------
We denote the position and velocity of the centre of mass of the $i$-th flapper
by $\gls{vxpi} = \left(\gls{vxpi_x},\gls{vxpi_y},\gls{vxpi_z}\right)$ and
$\gls{vupi} = \left(\gls{vupi_x},\gls{vupi_y},\gls{vupi_z}\right)$, respectively.
The superscripts $L\,(i=1)$ and $F\,(i=2)$ denote the leader and the
follower, respectively.
Translation in the lateral direction (\gls{vx_y}) and rotation about the
streamwise and vertical axes (\gls{vx_x} and \gls{vx_z}) are constrained.
The flappers are therefore free to translate in \gls{vx_x} and to rotate about a
spanwise axis through their centre of mass, \gls{vxpi}, with angle \gls{thi} and
angular velocity \gls{omi}.
Their vertical heaving motion is prescribed with sinusoidal laws
\begin{subequations} \label{eq:prob:motion}
\begin{align}
    \gls{vxpi_z}(t) &= \gls{A} \cos(2\pi \gls{f} t - \gls{phii} ) \,,
    \label{eq:prob:motion:pos}\\
    \gls{vupi_z}(t) &= - 2 \pi \gls{f} \gls{A} \sin(2\pi \gls{f} t - \gls{phii} ) \,,
    \label{eq:meth:motion:vel} %\\
\end{align}
\end{subequations}
where $\gls{A}$ is the heaving amplitude, $\gls{f}$ the frequency of 
oscillation and $\gls{phii}$ the phase shift.
Considering the above constraints and prescribed motion, the set of
Newton--Euler equations of motion is reduced to two scalar equations: the
linear momentum equation in the streamwise direction
\begin{equation}\label{eq:lin}
    \gls{rhop} \gls{Vp} \, \ddt{\gls{vupi_x}} =  \gls{vFi_x}\,,
\end{equation}
and the angular momentum equation in the spanwise direction
\begin{equation}\label{eq:mom}
    \gls{rhop} \gls{Vp} \gls{J}\, \ddt{\gls{omi}} = \gls{Mi} + \gls{Ki} \,,
\end{equation}
where \gls{vFi_x} is the \gls{vx_x}-component of the hydrodynamic force
$\gls{vFi}=\int_{S^{(i)}} \tau \cdot \bm{n} \, \mathrm{d}\sigma$
($\bm{n}$ represents the outward unit vector normal to the surface of the
flappers and $\tau = -\gls{p}\,\bm{I} + \gls{rhof}\gls{nu}(\nabla\gls{vu}
+ (\nabla\gls{vu})^{\top})$ the hydrodynamic stress tensor, with $\bm{I}$ the
identity tensor), and \gls{Mi} the \gls{vx_y}-component of the hydrodynamic 
torque
$\gls{vMi} = \int_{S^{(i)}} \bm{r}^{(i)} \times (\tau \cdot \bm{n}) \
\mathrm{d}\sigma$  ($\bm{r}^{(i)} = \bm{x}^{(i)} -
\gls{vxpi}$ is the position of a surface point relative to the centre of
mass).
Finally, \gls{Ki} is a restoring torque that results from the response of
a torsional spring to the pitching motion and it is described in more
detail below.
Note that gravity is not considered since the vertical motion of the flapper
is imposed.
% -----------------------------------------------------------------------------

% -----------------------------------------------------------------------------
We adopt the rotational spring model of \citet{schenk2026}, in which rotation
is considered around an axis aligned with the span ($\gls{vx_y}$) and passing 
through the centre of mass of the flapper, $\gls{vxpi}$.
The stiffness of the rotational spring, \gls{ki}, is constant; therefore, the
restoring torque varies linearly with the pitching angle
\begin{equation}
    \gls{Ki} = - \gls{ki} \gls{thi} \,.
    \label{eq:meth:K}
\end{equation}
% -----------------------------------------------------------------------------

% =============================================================================
%   PARAMETERS
% =============================================================================
For a fixed flapper geometry,
dimensional analysis reveals that the problem is governed by six
non-dimensional numbers.
Among the various possible combinations, we choose the density ratio between
the solid and the fluid,
\begin{equation}
\gls{rhor}=\frac{\gls{rhop}}{\gls{rhof}} \,,
\end{equation}
the Reynolds number based on the chord and the peak vertical velocity of the
centre of mass, $\gls{V} = 2\pi \gls{A} \gls{f}$,
\begin{equation}
\gls{ReV}= \frac{\gls{V}\gls{c}}{\gls{nu}} \,,
\end{equation}
the heaving amplitude normalised by the chord, $\gls{A}/\gls{c}$, the phase
shift difference between the follower and the leader,
\begin{equation} \label{eq:dphi}
    \gls{dphi} = \gls{phiF}-\gls{phiL},
\end{equation}
and the stiffness of each flapper in non-dimensional form,
\begin{equation}
\gls{kni} = \frac{\gls{ki}}{\gls{rhof}\gls{c;cb}\gls{V;sq}} \,.
\end{equation}
% -----------------------------------------------------------------------------

% -----------------------------------------------------------------------------
In order to compare our results with those of \gls{afg}, we set the density
ratio to $\gls{rhor}=10$, the Reynolds number to $\gls{ReV}=200$, and the
amplitude of the motion to $\gls{Aoc}=0.5$.
The phase difference is evaluated for
$\gls{dphi}= \{0^\circ,45^\circ,90^\circ,\ldots,315^\circ \}$.
Since we adopt the simplified elastic model of \citet{schenk2026}, rather
than the chordwise-flexible model of \gls{afg}, the stiffness is taken from
the former.
We denote by \gls{knOPT} the value of \gls{kn} that maximises the
propulsive speed of an isolated flapper identical to the one considered
here.
For $\gls{rhor}=10$, \citet{schenk2026} obtained
$\gls{knOPT}=0.28$.
% -----------------------------------------------------------------------------

% =============================================================================
%   METHODOLOGY
% =============================================================================
\section{Methodology}
\label{sec:meth}
% -----------------------------------------------------------------------------
We tackle the problem presented above by performing \glspl{dns}.
We use the in-house code TXPS, with which we solve the
Navier--Stokes equations of the incompressible flow \eqref{eq:ns}, together with
the Newton--Euler equations for the flappers \eqref{eq:lin} and
\eqref{eq:mom}, subject to the prescribed motion \eqref{eq:prob:motion}
and constraints described above.
The flow equations are discretised using second-order finite
differences on a staggered, uniform, Cartesian grid
and advanced in time with a semi-implicit
Adams--Bashforth/Crank--Nicolson scheme, treating the linear terms implicitly
and the nonlinear terms explicitly.
The presence of the flappers in the flow is modelled by the \gls{ibm} originally
proposed by \citet{uhlmann2005} and later extended in
\citet{moriche2021_settling} for non-spherical bodies.
% -----------------------------------------------------------------------------

% -----------------------------------------------------------------------------
In order to integrate over long time intervals, we follow the approach of
\citet{moriche2026}, which is well suited for problems where the body motion
along a non-periodic direction is not known beforehand, removing the necessity
for excessively large domains.
However, to apply the procedure presented in \citet{moriche2026}, we must draw
an analogy between the settling velocity and the Galileo number considered in
the reference, and the propulsive velocity and the flapping Reynolds number
used in this work, respectively.
One of the ingredients of the method is the definition of two reference frames:
i) the laboratory frame, in which the fluid is at rest in the absence of flapper
motion, and ii) a moving frame that translates with velocity $-U\gls{ex}$
relative to the laboratory frame.
The governing equations are solved in the moving frame and the value of $U$
is determined iteratively until it converges to the cycle-averaged propulsive
velocity of the flappers relative to the fluid within a prescribed tolerance.
However, because of the physical relevance of the laboratory frame, all results
presented in this work are referred to the laboratory frame.
% -----------------------------------------------------------------------------

% -----------------------------------------------------------------------------
The computational domain is a cuboid with side lengths
$[L_x,L_y,L_z]$, with periodicity imposed in the lateral and vertical
directions (\gls{vx_y} and \gls{vx_z}), as shown in Figure~\ref{fig:meth:sketch}b).
In the streamwise direction (\gls{vx_x}), we prescribe
$\gls{vu}=\left(U,0,0\right)$ at the inlet and a convective boundary condition at
the outlet.
To select the length of the domain in the streamwise direction, we have
considered both the distance from the inlet to the leader and from the follower
to the outlet.
We initialise the position of the leader so that its leading edge is $5\gls{c}$
downstream of the inlet and throughout the simulation we require this distance to
be in the range $[5\gls{c},6\gls{c}]$.
The follower is placed at an initial separation of $\gls{D0}=2\gls{c}$
from the leader (see Figure~\ref{fig:meth:sketch}a) and, based on the results of
\gls{afg}, we expect a maximum separation of $\gls{Dist}=3\gls{c}$ between the
flappers.
Therefore, we select a length $L_x=16\gls{c}$ in which we can accommodate the
vortices shed into the wake from the follower for a length of $6$--$7\gls{c}$
before they reach the outflow boundary.
In the lateral and vertical directions we set $L_y=L_z=8\gls{c}$, which is
somewhat larger than the domain used in \gls{afg}.
% -----------------------------------------------------------------------------

% -----------------------------------------------------------------------------
We select the spatial resolution based on the grid refinement study
presented in \citet{schenk2026} for a single flapper under otherwise
similar conditions, which identified $\gls{c}/\Delta x=64$ as an
appropriate resolution.
This results in a grid size of $[1024 \times 512 \times 512]$ points.
In addition to that, we select the time step  $\Delta t=\gls{T}/2048$
(with $\gls{T}=1/\gls{f}$), resulting in $CFL<0.3$ for all simulations.
% -----------------------------------------------------------------------------

% =============================================================================
%   AERODYNAMIC/PERFORMANCE COEFFICIENTS
% =============================================================================
\subsection{Definitions to evaluate the propulsive performance}
\label{sec:meth:performance}

This section introduces the performance indicators used to analyse the results
presented in this work.
To enable a direct comparison with \gls{afg}, we adopt the same definitions as
those used in their work.
All simulations reached a stable, periodic state with period $\gls{T}$, equal
to that of the imposed vertical motion.
In this state, the flappers maintained their equilibrium positions and their
mean separation remained constant.
Thus, a single oscillation cycle is sufficient to describe the solution,
and only cycle-averaging is needed to obtain mean quantities.
As a consequence, the cycle-averaged propulsive speed is the same for the
leader and the follower
\begin{equation}\label{eq:meth:velocity}
    \gls{Up} = \frac{-1}{\gls{T}}\int_{t^*}^{t^*+\gls{T}} \gls{vupL_x}(t)\,\wrt{t}
    = \frac{-1}{\gls{T}}\int_{t^*}^{t^*+\gls{T}} \gls{vupF_x}(t)\,\wrt{t}\,,
\end{equation}
where $t^*$ denotes an arbitrary time instant after convergence has been
reached and the minus sign is introduced for convenience when referring to
propulsive speeds with positive values.
Please recall that the quantities shown in this work are expressed in the
laboratory frame; therefore, there is no need to correct the expressions
in \eqref{eq:meth:velocity} with the inflow velocity imposed in the
simulations.
The mean distance between the leader and the follower is defined as
\begin{equation} \label{eq:meth:distance}
    \gls{D} = \frac{1}{\gls{T}}\int_{t^*}^{t^*+\gls{T}}  \left(
                 \gls{vxpF_x}(t)-\gls{vxpL_x}(t) \right)\,\wrt{t}\,,
\end{equation}
and the mean input power of each flapper
\begin{equation} \label{eq:meth:IP}
    \gls{Pi} = \frac{1}{\gls{T}}\int_{t^*}^{t^*+\gls{T}} \max(\gls{Pi_t}(t),\,0)\,\wrt{t}\,,
\end{equation}
where $\gls{Pi_t}=\gls{Ri}\cdot\gls{vupi_z}$, with $\gls{Ri}$ denoting the
vertical component of the actuator force acting at the centre of mass of the
$i$-th flapper.
The value of $\gls{Ri}$ can be readily obtained from the \gls{vx_z}-component of the
linear momentum equation
\begin{equation}
    \gls{Ri} = \gls{rhop}\gls{Vp} \ddt{\gls{vupi_z}(t)} - \gls{vFi_z} \,.
  \label{eq:meth:actuator}
\end{equation}
Please note that in the evaluation of the mean input power, negative values of
\gls{Pi_t} are discarded, reflecting the assumption that the actuator is unable
to recover energy, in line with previous studies~\citep{berman2007,michelin2009}.
Furthermore, although the torsional spring can passively store and release
energy through the rotational degree of freedom, the present definition of
\gls{Pi} only accounts for the active heaving actuation and does not include
any form of energy recovery.
Finally, if we use the kinetic energy based on the mean propulsive speed as a
measure of the energy delivered by the system, and the combination of the
mean input power from the leader and follower as its energy demand, we can
combine them to evaluate the efficiency of the system in terms of energy
\begin{equation}
   \gls{E} = \frac{2 \cdot \frac{1}{2}\gls{rhop}\gls{Vp}\gls{Up;sq}}
    {\left(\gls{PL}+\gls{PF}\right) \gls{T}}
    = \frac{\gls{rhop}\gls{Vp}\gls{Up;sq}}
    {\left(\gls{PL}+\gls{PF}\right) \gls{T}} \,.
    \label{eq:meth:efficiency}
\end{equation}
This work-based definition is particularly suited to cases where the thrust/drag
force is not directly available
\citep[for more details see][their section 2.5.5]{zhang2022}.
% -----------------------------------------------------------------------------

% -----------------------------------------------------------------------------
Now let us focus on how to evaluate the gain/loss of performance of the tandem
configuration with respect to the case of a single flapper.
The results for an isolated flapper are indicated with the superscript $S$ and
the values presented here are taken from \citet{schenk2026}.
We first define the velocity ratio between the tandem and the isolated flapper
\begin{equation}
    \gls{UpR} = \frac{\gls{Up}}{\gls{UpS}} \,. \label{eq:meth:UpRatio}
\end{equation}
Second, we define the ratio of the power input of the $i$-th flapper relative to
an isolated flapper
\begin{equation}
    \gls{Pir} = \frac{\gls{Pi}}{\gls{PS}} \,. \label{eq:meth:IPratio}
\end{equation}
Finally, we compute the ratio of propulsive efficiency between the tandem
configuration and the single flapper
\begin{equation}
   \gls{Er} = \frac{\gls{E}}{\gls{ES}} \,, \label{eq:meth:Eratio}
\end{equation}
where the definition of \gls{ES} is included for clarity
\begin{equation}
   \gls{ES} = \frac{\frac{1}{2}\gls{rhop}\gls{Vp}\gls{UpS;sq}}
    {\gls{PS} \gls{T}} \,.
    \label{eq:meth:efficiencySingle}
\end{equation}
% -----------------------------------------------------------------------------

% =============================================================================
% =============================================================================
%   RESULTS
% =============================================================================
% =============================================================================
\section{Results} \label{sec:results}

\subsection{Emergent dynamics of the tandem configuration}
\label{sec:tandem}

% -----------------------------------------------------------------------------
We first consider the tandem configuration in which both the leader and
the follower have the same torsional spring stiffness, $\gls{knL} = \gls{knF}
= \gls{knOPT}$.
Here, $\gls{knOPT}=0.28$ is the normalised stiffness
that maximises the propulsive speed of the corresponding isolated flapper
\citep{schenk2026}.
% -----------------------------------------------------------------------------

% =============================================================================
%   FLOW VISUALISATION
% =============================================================================
\begin{figure}[b!]
  \centering
  \includegraphics[width=\textwidth]{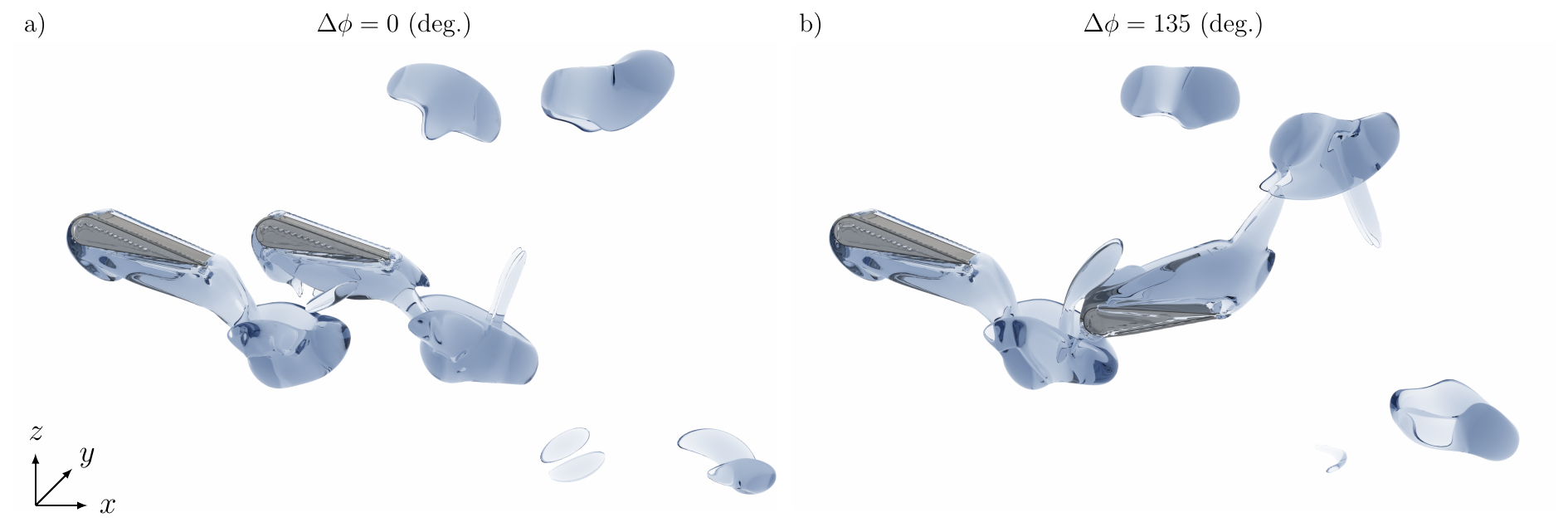}
  \caption{Flow visualisation around the flappers in tandem formation with
    the same spring stiffness at $t/\gls{T}=0.78125$.
    Flow is visualised in terms of iso-surfaces of the $\gls{Q}$-criterion for
    $\gls{Q} \, \gls{c;sq}/\gls{V;sq}=0.5$: (a) $\gls{dphi}=0^\circ$ and
    (b) $\gls{dphi}=135^\circ$.}
  \label{fig:res:flowvis}
\end{figure}
% -----------------------------------------------------------------------------
Figure~\ref{fig:res:flowvis} illustrates the flow around the flappers
for $\gls{dphi}=0^\circ$ and $\gls{dphi}=135^\circ$ at $t/\gls{T}=0.78125$,
by means of iso-surfaces of the $\gls{Q}$-criterion.
The leader sheds a vortex ring during each stroke, and the rings move away
from it along oblique paths due to their self-induced velocity.
As a result, the wake splits into two rows above and below the path of the
flappers, as reported for low-aspect-ratio hydrofoils by
\citet{dong2006} and \citet{kurt2020}.
The follower swims inside this wake, so that the vortex structures shed by
the leader reach it before they are convected further downstream.
The position of the vortex rings relative to the follower at this instant
depends on \gls{dphi} and does not reflect the strength of the interaction.
For $\gls{dphi}=135^\circ$ the follower is located further downstream than
for $\gls{dphi}=0^\circ$, which anticipates the dependence of the distance
between the flappers on \gls{dphi} discussed below.
% -----------------------------------------------------------------------------

% =============================================================================
%   EQUILIBRIUM DISTANCE
% =============================================================================
\begin{figure}[t!]
    \centering
    \includegraphics[width=0.98\textwidth]{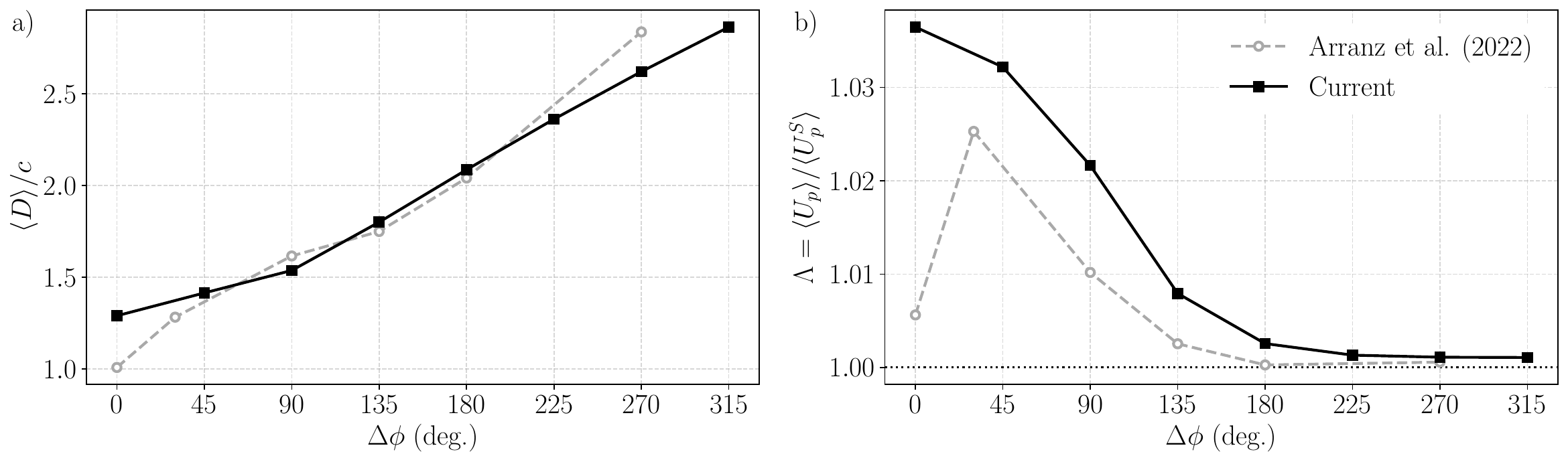}
    \caption{%
    (a) Mean distance between the two flappers ($\gls{D}/\gls{c}$) versus
    phase shift of the motion ($\gls{dphi}$).
    The inset shows the instantaneous pressure field
    ($\gls{p}/(\gls{rhof} \gls{V;sq})$) around the two flappers at
    $\gls{dphi}=0^\circ$ during the stroke reversal.
    (b) Streamwise propulsive-velocity ratio of the tandem configuration
    normalised by the isolated-flapper value ($\gls{UpR}=\gls{Up}/\gls{UpS}$).
    The results from \gls{afg} are included for comparison.
    }
    \label{fig:res:tandem_no:distvel}
\end{figure}

% -----------------------------------------------------------------------------
The flappers are initially placed such that their leading edges are
$\gls{D0}=2\gls{c}$ apart.
The equilibrium distance may depend on the initial separation.
For example, \gls{afg}  found that the case with zero phase shift converged to 
two different equilibrium positions depending on the initial condition.
The equilibria reported here are those reached from $\gls{D0}=2\gls{c}$.
Figure~\ref{fig:res:tandem_no:distvel}a) shows the mean distance between the
flappers as a function of the phase shift.
The distance increases approximately linearly with the phase shift,
with a small deviation
in the slope for the cases with $\gls{dphi}\le 90^\circ$.
This trend is consistent with the reference case of \gls{afg} and with previous
 results for two-dimensional configurations
\citep{lin2019,newbolt2019,ryu2020}.
The vortices shed by the leader are advected downstream with a fixed spacing,
set by the wake wavelength $\lambda_l$.
The follower settles at a location where
its flapping cycle stays favourably phased with these wake structures.
Increasing $\gls{dphi}$ shifts this preferred position downstream by a
corresponding fraction of $\lambda_l$, as proposed by
\citet{portugal2014}, which produces the approximately linear dependence of
the distance on the phase difference \citep{ramananarivo2016,newbolt2019}.
% -----------------------------------------------------------------------------

% -----------------------------------------------------------------------------
The present results differ from those of \gls{afg} mainly at small phase
differences ($\gls{dphi}<90^\circ$), where the present model predicts a
larger distance.
Both models agree closely for $90^\circ\le\gls{dphi}\le225^\circ$,
whereas at $\gls{dphi}=270^\circ$ the present distance is
about $0.2\gls{c}$ smaller.
The difference at small phase shifts does not result from a geometric constraint,
since the minimum gap between the trailing edge of the leader and the leading
edge of the follower remains above $0.29\gls{c}$ even for $\gls{dphi}=0$, where
the equilibrium distance is smallest.
Instead, the inset of Figure~\ref{fig:res:tandem_no:distvel}a) shows a region of
high pressure around the leading edge of the follower at stroke reversal, which
acts against its forward motion.
The flexible flappers of \gls{afg} are flat and clamped at the leading edge,
and are therefore expected to be less susceptible to this pressure effect.
This difference in geometry may explain the smaller equilibrium distance obtained
by \gls{afg} at small phase shifts.
% -----------------------------------------------------------------------------

% =============================================================================
%   VELOCITY RATIO
% =============================================================================
In Figure~\ref{fig:res:tandem_no:distvel}b), we compare the propulsive
speed of the tandem with that of the isolated flapper through the ratio
$\gls{UpR}=\gls{Up}/\gls{UpS}$.
The maximum gain is approximately $3.65\%$ at $\gls{dphi}=0^\circ$,
compared with approximately $2.5\%$ in \gls{afg}.
Beyond the maximum of \gls{afg}, both data sets show a speed-up that
decreases with increasing \gls{dphi}.
This small enhancement lies well below the speed-ups reported for
two-dimensional viscous simulations at comparable Reynolds numbers:
\citet{lin2019} report up to $50\%$, \citet{ryu2020} up to about $47\%$, and
\citet{peng2018} up to $19\%$.
We relate the lower enhancement in three dimensions mainly to the
differences in vortex dynamics, namely the decay of the vortex
intensity and the vortex topology: vortex rings in 3D versus
infinitely long spanwise rolls in 2D.
Furthermore, the vortices shed in the reverse von K\'arm\'an street
in the 2D case stay close to the centreline of the wake, keeping
the associated momentum aligned with the follower, whereas in a
3D problem the vortex rings spread laterally.
These two aspects reduce the momentum transferred to the follower, while
the flexible plates of \citet{peng2018} and \citet{ryu2020} may also
contribute to their larger speed-ups.
In experiments at much higher Reynolds numbers, \citet{ramananarivo2016}
measured speed-ups of up to $25\%$.
In similar experiments, \citet{newbolt2019} found the speed of the leader to
be nearly independent of the motion of the follower, within $4\%$.
Using two-dimensional vortex-sheet simulations with a skin-friction model,
\citet{fang2026} found a pair of heaving rigid plates with identical
kinematics to be always slower than an isolated plate.
These results show that the sign and magnitude of the speed-up depend on
the Reynolds number, the kinematics and the structural model.
% -----------------------------------------------------------------------------

% =============================================================================
%   INPUT POWER RATIO
% =============================================================================
\begin{figure}[t!]
    \centering
    \includegraphics[width=\textwidth]{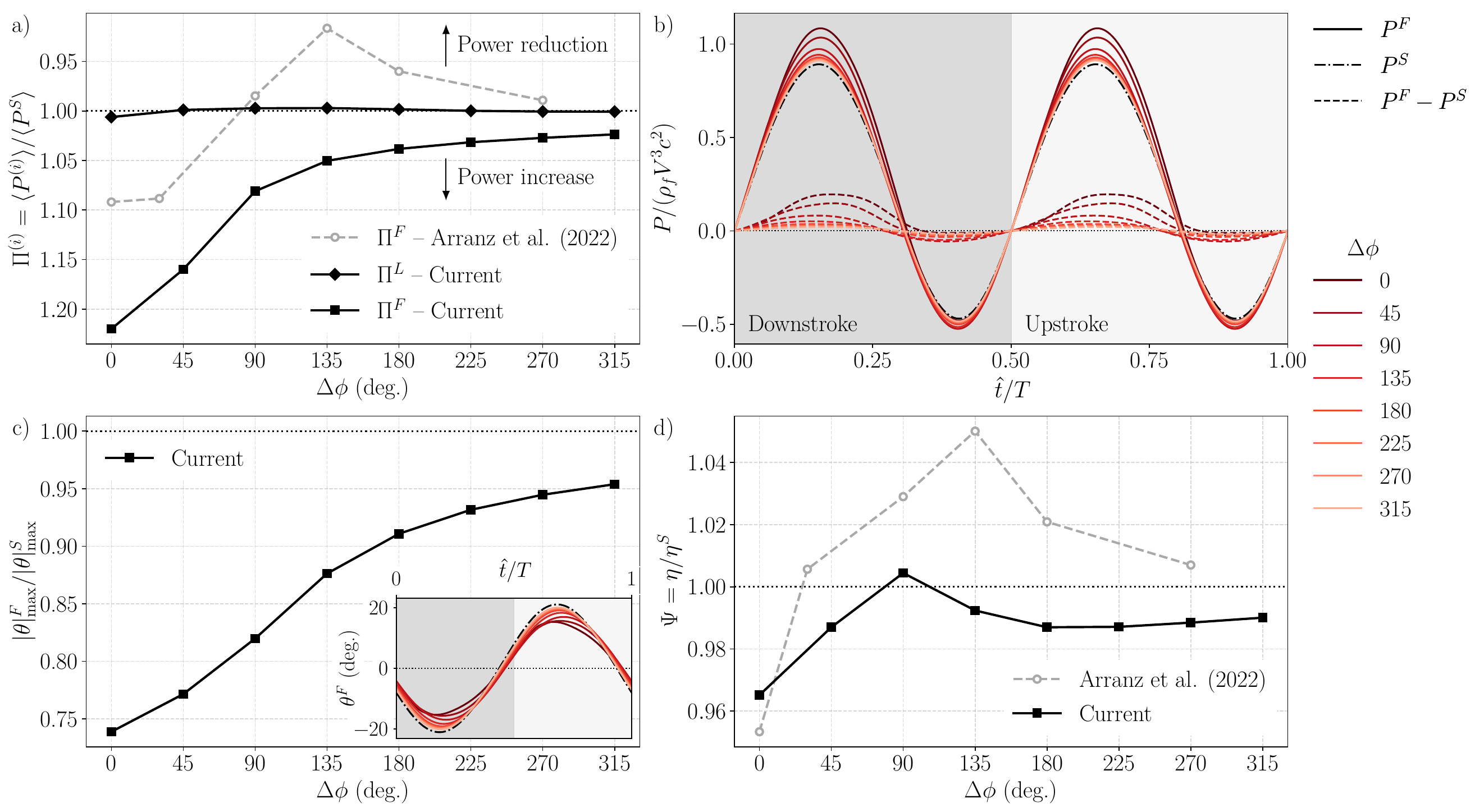}
    \caption{%i
    (a) Cycle-averaged input-power ratio of the leader (\gls{PiL}) and follower
    (\gls{PiF}) as a function of $\gls{dphi}$; the vertical axis is inverted so
    that reductions in input power lie above $\gls{Pir}=1$.
    (b) Instantaneous follower input power $\gls{Pi_F}$ over one cycle, shifted
    by $\gls{dphi}$, together with the isolated flapper $\gls{Pi_S}$
    (dash-dotted) and the difference $\gls{Pi_F}-\gls{Pi_S}$ (dashed).
    (c) Ratio of the maximum pitching angle of the follower to that of the
    isolated flapper, $\gls{thFA}/\gls{thSA}$; the inset shows the follower
    pitching angle $\gls{thF}$ over one cycle.
    (d) Propulsive-efficiency ratio $\gls{Er}$.
    Colours denote $\gls{dphi}$, and grey symbols the results of \gls{afg}.
    }
    \label{fig:res:tandem_no:quat}
\end{figure}

% -----------------------------------------------------------------------------
Figure~\ref{fig:res:tandem_no:quat}a) shows the ratio between the input power of
each flapper in the tandem configuration and that of the isolated flapper,
\gls{Pir} (see Eq.~\eqref{eq:meth:IPratio}), as a function of \gls{dphi}.
Values smaller than unity ($\gls{Pir} < 1$) represent a gain in performance,
since less energy is consumed by the system.
The input power of the leader is essentially the same as for the isolated
flapper ($\gls{PiL} \approx 1$) for all cases.
The follower, in contrast, always requires more power than the isolated
flapper ($\gls{PiF} > 1$).
The penalty is largest for synchronous motion ($\gls{dphi} = 0$), where the
follower requires about $22\%$ more power.
With increasing \gls{dphi}, the excess input power of the follower decreases
monotonically, falling below $3\%$ at $\gls{dphi} = 315^\circ$.
\gls{afg}, conversely, report power savings for phase shift values
$\gls{dphi}\ge 90^\circ$, with a maximum saving of $7.5\%$ at
$\gls{dphi}=135^\circ$, beyond which the savings decrease monotonically with
increasing \gls{dphi}.
As in the present results, \gls{afg} also find a power penalty, although less
pronounced, when the gap between the trailing edge of the leader and the leading
edge of the follower falls below half a chord ($\gls{dphi}\le45^\circ$).

Figure~\ref{fig:res:tandem_no:quat}b) shows the instantaneous input power of the
follower $\gls{Pi_F}$ over one cycle  together with that of the isolated flapper
\citep{schenk2026}.
Since the solutions obtained here are fully periodic, no phase averaging is
performed.
To allow a direct comparison, we define the shifted time
$\gls{tshift} = t - \gls{dphi}/(2\pi \gls{f})$, which aligns the follower's
downstroke/upstroke with that of the isolated flapper.
The follower closely follows the isolated curve over most of the cycle, but
exceeds it during the two power-input peaks, where the actuator does positive
work to drive the heaving motion.
The difference is largest when the motion of both flappers is in phase
($\gls{dphi}=0$) and decreases steadily with increasing \gls{dphi}, consistent
with Figure~\ref{fig:res:tandem_no:quat}a).
Since the mean input power retains only positive values of \gls{Pi_t}
(Eq.~\ref{eq:meth:IP}), the energy returned during deceleration is not
accounted for.
The higher mean power of the follower therefore stems predominantly from
these peaks.

We relate this penalty to the pitching motion of the follower.
Figure~\ref{fig:res:tandem_no:quat}c) displays the ratio of the
maximum pitching angle of the follower to that of the isolated flapper,
$\gls{thFA}/\gls{thSA}$.
This ratio stays below unity for all values of \gls{dphi} and increases
monotonically with \gls{dphi}, approaching the isolated value.
The inset of Figure~\ref{fig:res:tandem_no:quat}c) shows that, for all
values of \gls{dphi}, the follower reaches its maximum, minimum and zero
pitching angles at nearly the same instants of the cycle as the isolated
flapper.
The wake of the leader thus mainly reduces the pitching amplitude of the
follower, without shifting its phase.
The largest reduction in pitching amplitude coincides with the largest
power penalty, when the follower swims closest to the leader.
Both flappers have the stiffness that maximises the propulsive speed of
an isolated flapper.
The pitching amplitude of the leader, however, remains close to the
value of the isolated flapper (not shown).
This suggests that the follower operates away from the condition that
maximises its propulsive performance in isolation, although the interaction
between the two flappers complicates this comparison.
For an isolated flapper, \citet{schenk2026} show that reducing the pitching
amplitude below its optimal value increases the input power required to heave
the flapper.
This is consistent with the power penalty observed for the follower.
In contrast, the chordwise-flexible flapper considered by \gls{afg} can
adapt its deformation to the induced flow.
Moreover, \gls{afg} impose the heaving motion at the leading edge, so that
the plate, whose centre of mass lies downstream of the leading edge, is
also bent by its own inertia.
The present flapper, in contrast, pitches about its centre of mass, where the
imposed heave exerts no moment, so that only the hydrodynamic and spring
moments drive its rotation.
These differences may explain why the power savings reported by \gls{afg} for
$\gls{dphi}\ge 90^\circ$ are not recovered by the present model.
% -----------------------------------------------------------------------------

% =============================================================================
%   PROPULSIVE EFFICIENCY RATIO
% =============================================================================
Finally, Figure~\ref{fig:res:tandem_no:quat}d) shows the propulsive-efficiency
ratio \gls{Er} (see Eq.~\ref{eq:meth:Eratio}) as a function of the phase
shift $\gls{dphi}$.
From Eqs.~\eqref{eq:meth:efficiency}--\eqref{eq:meth:efficiencySingle},
$\gls{Er}=2\gls{UpR}^2/(\gls{PiL}+\gls{PiF})$, so that $\gls{Er}$ combines
the gain in propulsive speed with the input power of both flappers.
For the present model, \gls{Er} increases from its in-phase value to a
maximum of approximately $1.005$ at $\gls{dphi}=90^\circ$, before dropping
slightly below unity for larger phase shifts.
Apart from this marginal gain of about $0.5\%$ at $\gls{dphi}=90^\circ$,
the tandem shows no efficiency benefit with respect to the isolated flapper.
This differs clearly from \gls{afg}, who obtained values
above unity for all phase shifts  except $\gls{dphi}=0$.
Therefore, in the present configuration, the marginal gain in propulsive
speed does not compensate for the additional power demanded by the follower.
Since the leader is essentially unaffected in terms of input power,
the stiffness of the follower is a natural parameter to improve
the performance of the tandem, as examined below.
% -----------------------------------------------------------------------------

% =============================================================================
% -----------------------------------------------------------------------------
%   FOLLOWER STIFFNESS
% -----------------------------------------------------------------------------
% =============================================================================
\subsection{Effect of the follower stiffness on the tandem performance}
\label{sec:res:optimisation}

% -----------------------------------------------------------------------------
It was shown in the previous section, for flappers with the same torsional
stiffness, that the leader is largely unaffected by the follower in the parameter
space considered here.
In this section we therefore maintain $\gls{knL}=\gls{knOPT}$, and vary $\gls{knF}$
independently.
The rationale behind this is that the environment in which the follower operates
is markedly different from that of the leader.
The leader encounters quiescent fluid, whereas the follower swims through the
vortical wake of the leader.
We characterise the new configuration by the stiffness ratio between the follower
and the leader
\begin{equation}
    \gls{knr} = \frac{\gls{knF}}{\gls{knL}} = \frac{\gls{knF}}{\gls{knOPT}}.
\end{equation}
% -----------------------------------------------------------------------------

% -----------------------------------------------------------------------------
For each of the eight phase differences evaluated $\gls{dphi} \in
\{0^\circ,45^\circ,\dots,315^\circ\}$, we perform four additional simulations with
\begin{equation}
    \gls{knr} =
    \begin{cases}
        0.8,\ 0.9 & \gls{knr} < 1 \ \rightarrow \ \text{softer follower} \\
        1.1,\ 1.2 & \gls{knr} > 1 \ \rightarrow \ \text{stiffer follower}
    \end{cases}
\end{equation}
%
% -----------------------------------------------------------------------------
Preliminary simulations at a lower resolution indicate that, for
$\gls{knr}\lesssim 0.7$, the pitching amplitude of the follower becomes so large
that its drag exceeds its thrust.
The follower then falls behind the leader, and the equilibrium distance increases
rapidly.
These configurations cannot be accommodated in the present domain
of length $L_x=16\gls{c}$ and are therefore excluded from the analysis.
% -----------------------------------------------------------------------------

% =============================================================================
%   EQUILIBRIUM DISTANCE
% =============================================================================
\begin{figure}[t]
  \centering
  \includegraphics[width=0.95\textwidth]{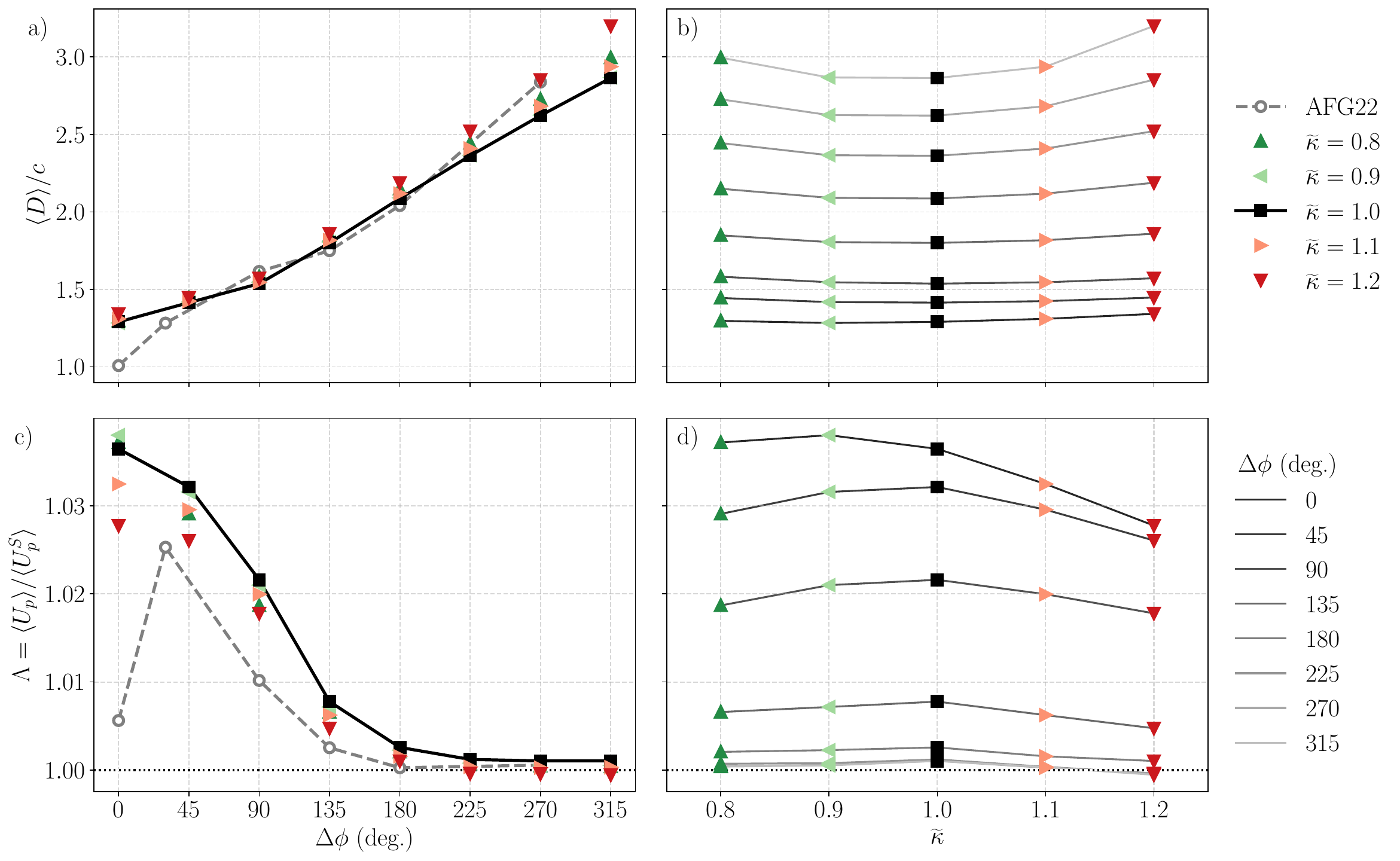}
  \caption{(a,b) Cycle-averaged equilibrium distance $\gls{D}/\gls{c}$
   between the two flappers and (c,d) propulsive-velocity ratio $\gls{UpR}$
   of the tandem configuration.
   Both quantities are represented against the phase shift \gls{dphi}
   (left column) and the stiffness ratio \gls{knr} (right column).
   The baseline case $\gls{knr}=1$ corresponds to identical flappers, and
   the results from \gls{afg} are included in (a,c) for
   comparison.}
  \label{fig:res:tandem_opt:distance}
\end{figure}

% -----------------------------------------------------------------------------
Figure~\ref{fig:res:tandem_opt:distance} shows the effect of varying
the follower's stiffness on the mean distance between the flappers and
on the propulsive-velocity ratio.
For softer followers ($\gls{knr} < 1$), the mean distance is
roughly the same as in the baseline cases ($\gls{knr}=1$) discussed above.
For stiffer followers ($\gls{knr} > 1$) the mean distance increases
slightly, and this increase grows with \gls{dphi}.
Nevertheless, the approximately linear behaviour of the mean distance with respect
to the phase shift is retained.
Interestingly, even the deviation for low values of $\gls{dphi}\le90^\circ$ is
reproduced.
This is consistent with the follower locking onto quantised positions set by
$\gls{dphi}$ and the wavelength of the leader's wake, rather than by its own
stiffness \citep{becker2015,ramananarivo2016,newbolt2019}.
The present results extend the linear relation between \gls{dphi} and the
equilibrium distance, reported by \citet{kanso2026} for rigid and flexible
swimmers in two and three dimensions, to flappers of unequal stiffness.
% -----------------------------------------------------------------------------

% -----------------------------------------------------------------------------
Similarly, the propulsive-speed ratio depends only weakly on \gls{knr},
Figure~\ref{fig:res:tandem_opt:distance}c,d).
Softer followers leave \gls{UpR} essentially unchanged at all phase differences.
Stiffer followers reduce \gls{UpR} mainly at small phase differences,
where the interaction between the flappers is strongest.
Even at $\gls{dphi}=0^\circ$, the difference between the softest and the stiffest
follower remains below 1\%.
As for the tandem with equal stiffness (\S~\ref{sec:tandem}), the speed-up
therefore remains at a few per cent.
Together, these results indicate that the equilibrium state of the tandem
is governed primarily by $\gls{dphi}$ rather than by $\gls{knr}$.
% -----------------------------------------------------------------------------

% =============================================================================
%   INPUT POWER AND EFFICIENCY
% =============================================================================
\begin{figure}[t]
   \centering
   \includegraphics[width=0.9\textwidth]{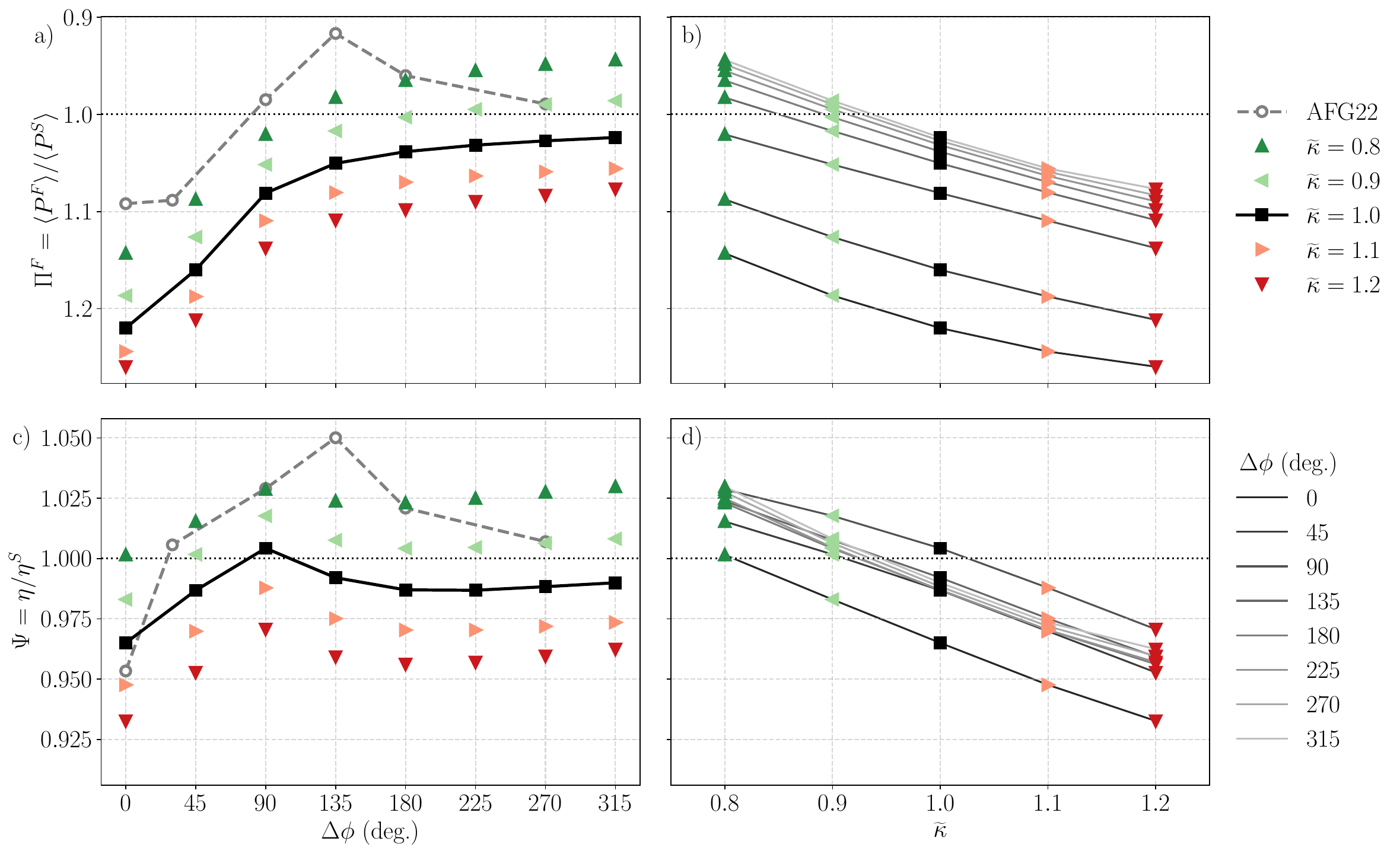}
   \caption{(a,b) Input-power ratio \gls{PiF} of the follower and (c,d)
      propulsive-efficiency ratio \gls{Er} of the tandem configuration.
      Both quantities are represented against the phase shift \gls{dphi}
      (left column) and the stiffness ratio \gls{knr} (right column).
      The vertical axis in (a,b) is inverted, and the results from
      \gls{afg} are included for comparison.
   }
   \label{fig:res:tandem_opt:power}
\end{figure}

%------------------------------------------------------------------------------
We now turn to the  input-power ratio of the follower \gls{PiF}, which
accounted for the main performance loss of the baseline cases
in \S~\ref{sec:tandem}.
The values for the leader are omitted, since $\gls{PiL} \approx 1$
for all cases evaluated.
Interestingly, we observe a consistent reduction in required input power for
softer followers, and an increase for stiffer followers
(Figure~\ref{fig:res:tandem_opt:power}a,b).
Furthermore, \gls{PiF} varies approximately linearly with \gls{knr}, as
seen in Figure~\ref{fig:res:tandem_opt:power}b).
As a consequence, \gls{knr} shifts \gls{PiF} almost uniformly, while its
dependence on \gls{dphi} remains that of the baseline case,
Figure~\ref{fig:res:tandem_opt:power}a).
Relative to the baseline, the input power of the follower decreases
by approximately 8\% for $\gls{knr}=0.8$ and increases by about
6\% for $\gls{knr}=1.2$.
An isolated flapper shows the same trend: a softer spring increases its
pitching amplitude and reduces the input power required to heave it
\citep{schenk2026}.
For $\gls{knr}=0.8$, this reduction brings $\gls{PiF}$ below unity for
$\gls{dphi}\ge 135^\circ$, so that the follower requires less power
than an isolated flapper.
For $\gls{knr}=0.9$, $\gls{PiF}$ falls below unity only at the largest phase
differences.
Thus, tuning the follower's stiffness offers a passive way to reduce the
required input power of the tandem configuration relative to an isolated
flapper.

% -----------------------------------------------------------------------------
Figures~\ref{fig:res:tandem_opt:power}c,d) show the propulsive-efficiency
ratio \gls{Er} as a function of \gls{dphi} and \gls{knr}.
\gls{Er} accounts for the propulsive speed and for the input power of both
flappers.
The improvement for softer followers, and the deterioration
for stiffer ones, are retained in this integral measure.
All softer followers show an improvement with respect to the isolated case,
except for $\gls{knr}=0.9$ at $\gls{dphi}=0$.
For $\gls{knr}=0.8$, $\ \gls{Er}>1$ at every phase difference, although the 
margin is only about 0.1\% at $\gls{dphi}=0^\circ$.
An efficiency gain of a trailing body relative to an isolated one
has been widely reported for two-dimensional configurations.
\citet{boschitsch2014} found higher efficiency for in-phase flapping, and
\citet{park2018} reported an efficiency benefit for a trailing flexible fin.
Similarly, \citet{dai2018} found a collective cost of transport up to $16\%$
lower than that of isolated elastic plates.
The magnitude of this gain, however, depends strongly on the configuration.
For example, for a side-by-side pair of pitching foils, \citet{ormonde2024}
reported a collective efficiency gain of up to $9\%$ when self-propelled,
compared with up to $40\%$ when held at a fixed position.
The gain is generally smaller in three dimensions, although \gls{afg} reported
similarly favourable efficiency gains for their chordwise-flexible flapper in
a tandem configuration.
The present results show that a comparable gain can be achieved with a rigid
flapper and a torsional spring by tuning the follower stiffness.
% -----------------------------------------------------------------------------

% =============================================================================
%   INPUT POWER / PITCHING PHASE PLOTS
% =============================================================================
\begin{figure}[t!]
  \centering
  \includegraphics[width=0.9\textwidth]{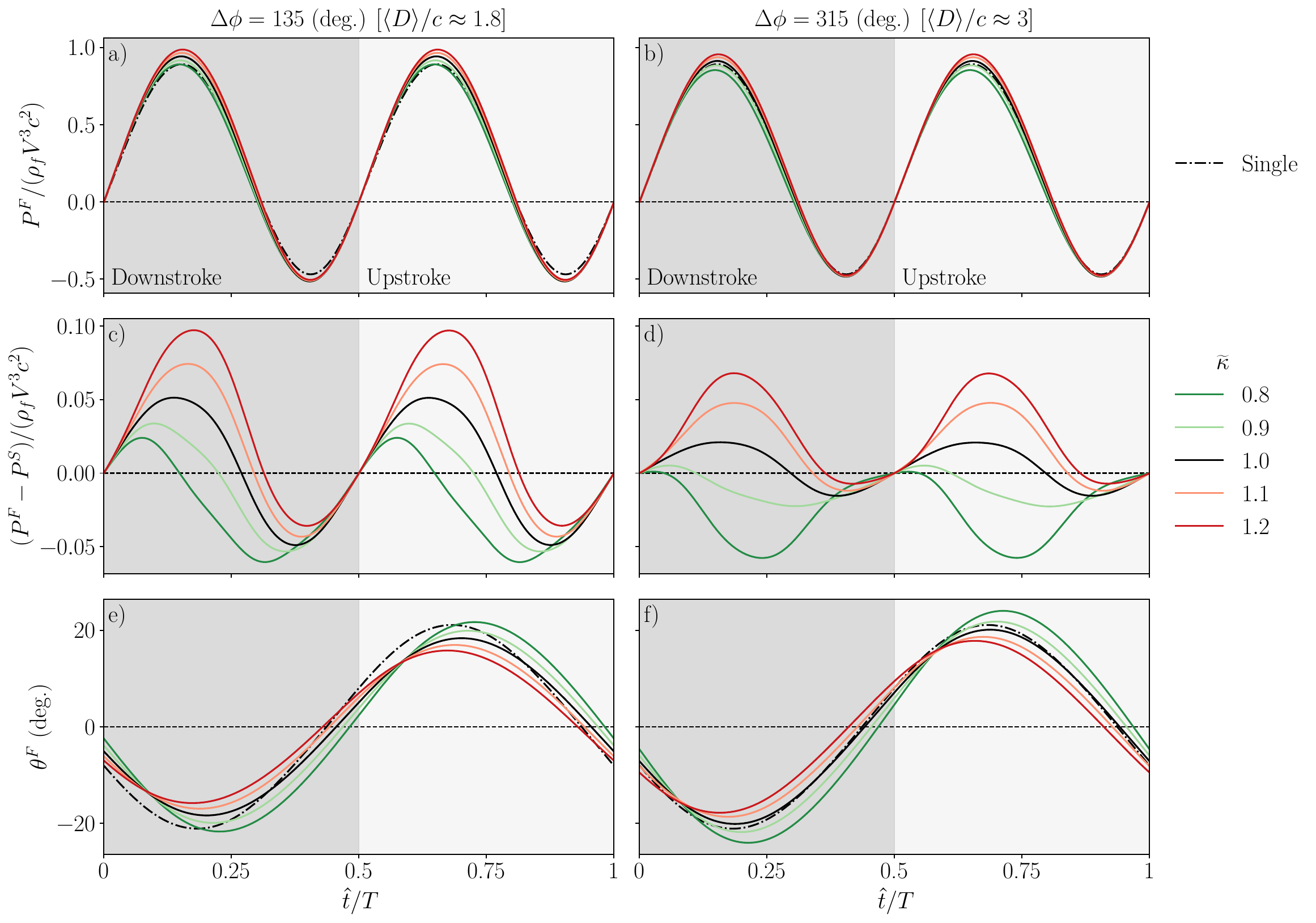}
    \caption{(a,b) Instantaneous input power of the follower \gls{Pi_F} and
    (c,d) its difference with respect to the isolated case
    ($\gls{Pi_F}-\gls{Pi_S}$) during one cycle, together with
    (e,f) the pitching angle of the follower \gls{thF}.
    The cases shown correspond to $\gls{dphi}=135^\circ$ (left column) and
    $315^\circ$ (right column), for all the stiffness ratios evaluated,
    $\gls{knr}=\{0.8,0.9,1,1.1,1.2\}$.
  }
  \label{fig:res:tandem_opt:power_phase}
\end{figure}

% -----------------------------------------------------------------------------
\def\myfig{\ref{fig:res:tandem_opt:power_phase}}
To identify the origin of the power reduction,
Figure~\ref{fig:res:tandem_opt:power_phase} shows the instantaneous input
power of the follower for all five stiffness ratios at
$\gls{dphi}=135^\circ$ and $\gls{dphi}=315^\circ$.
At $\gls{dphi}=135^\circ$, $\gls{knr}=0.8$ is the only stiffness ratio for
which $\gls{PiF}<1$, so that this phase difference marks the transition to
power saving.
At $\gls{dphi}=315^\circ$, $\gls{knr}=0.8$ yields the lowest input power of
all cases considered.
The evolution of \gls{Pi_F} shown in Figures~\myfig{}a,b) resembles
that of the baseline case shown in Figure~\ref{fig:res:tandem_no:quat}b),
with small differences observed near the positive peaks.
The negative peaks of \gls{Pi_F}, however, vary little with \gls{knr},
and do not contribute to any of the efficiency measures (Eq. \ref{eq:meth:IP}).
The differences $\gls{Pi_F} - \gls{Pi_S}$ displayed in Figure~\myfig{}c,d)
show that the ordering of the stiffness ratios holds almost instantaneously,
except near the end of each stroke at $\gls{dphi}=315^\circ$.
At $\gls{dphi}=135^\circ$ and $\gls{knr}=0.8$, \gls{Pi_F} exceeds \gls{Pi_S}
during roughly the first 20\% of each stroke and is lower thereafter,
Figure~\ref{fig:res:tandem_opt:power_phase}c).
Only the part of this reduction that occurs while $\gls{Pi_F}>0$, i.e.\
before $\gls{tshift}/\gls{T}\approx 0.35$ in the downstroke, enters \gls{PF}.
For $\gls{knr}=0.8$, $\gls{Pi_F} - \gls{Pi_S}$ changes sign earlier than for
the stiffer followers, so that a larger part of the reduction falls within this
interval.
This yields a net reduction in the required input power.
At $\gls{dphi}=315^\circ$ and
$\gls{knr}=0.8$, \gls{Pi_F} remains below \gls{Pi_S} throughout the cycle.
% -----------------------------------------------------------------------------

% -----------------------------------------------------------------------------
Figures~\myfig{}e,f) show the pitching angle of the follower $\gls{thF}$ (or
geometric angle of attack).
The softer the follower, the larger its pitching amplitude, since the weaker
torsional spring restrains it less.
The amplitude \gls{thFA} ranges from approximately $16^\circ$ for
$\gls{knr}=1.2$ to approximately $24^\circ$ for $\gls{knr}=0.8$.
An isolated flapper responds to a softer spring in the same way, since the
hydrodynamic moment that damps the rotation gains importance relative to the
spring moment \citep{schenk2026}.
At the same time, the peak of \gls{thF} is delayed slightly and
monotonically as the follower becomes softer.
The phase shift acts differently (Figure~\ref{fig:res:tandem_no:quat}c inset).
It changes the pitching amplitude but leaves its timing nearly unchanged,
and \gls{thFA} remains below \gls{thSA} for all phase shifts.
Varying $\gls{knr}$, in contrast, shifts the peak in time and allows
\gls{thFA} to exceed $\gls{thSA}\approx 21.1^\circ$.
Among the cases in Figure~\myfig{}, this occurs for $\gls{knr}=0.8$ at
$\gls{dphi}=135^\circ$, and for $\gls{knr}=0.8$ and $0.9$ at
$\gls{dphi}=315^\circ$.
These three cases are also the only ones in the figure with $\gls{PiF}<1$,
i.e.\ with a follower that requires less input power than the isolated flapper.
The same link between a larger pitching amplitude and a lower input power
holds for the isolated flapper \citep{schenk2026}.
A softer follower thus reverses the trend found for equal stiffness, where the
wake of the leader reduces the pitching amplitude of the follower and
increases $\gls{PiF}$.
Since the leader pitches with nearly the amplitude of the isolated flapper
(\S~\ref{sec:tandem}), these cases correspond to a follower pitching with a
larger amplitude than the leader.
Because \gls{Up} is nearly constant for all stiffness ratios, the lower
\gls{PiF} also raises \gls{Er}.
A higher collective efficiency for a follower pitching with a larger amplitude
than the leader has also been reported for purely pitching foils held at a
fixed position \citep{kurt2021}.
For a self-propelled tandem of purely pitching foils, in contrast,
\citet{han2025} obtained a higher efficiency with the opposite mismatch,
a leader pitching with a larger amplitude than the follower.
% -----------------------------------------------------------------------------

% =============================================================================
%   VISUALIZATION
% =============================================================================
\begin{figure}[!t]
  \centering
  \includegraphics[width=1.0\textwidth]{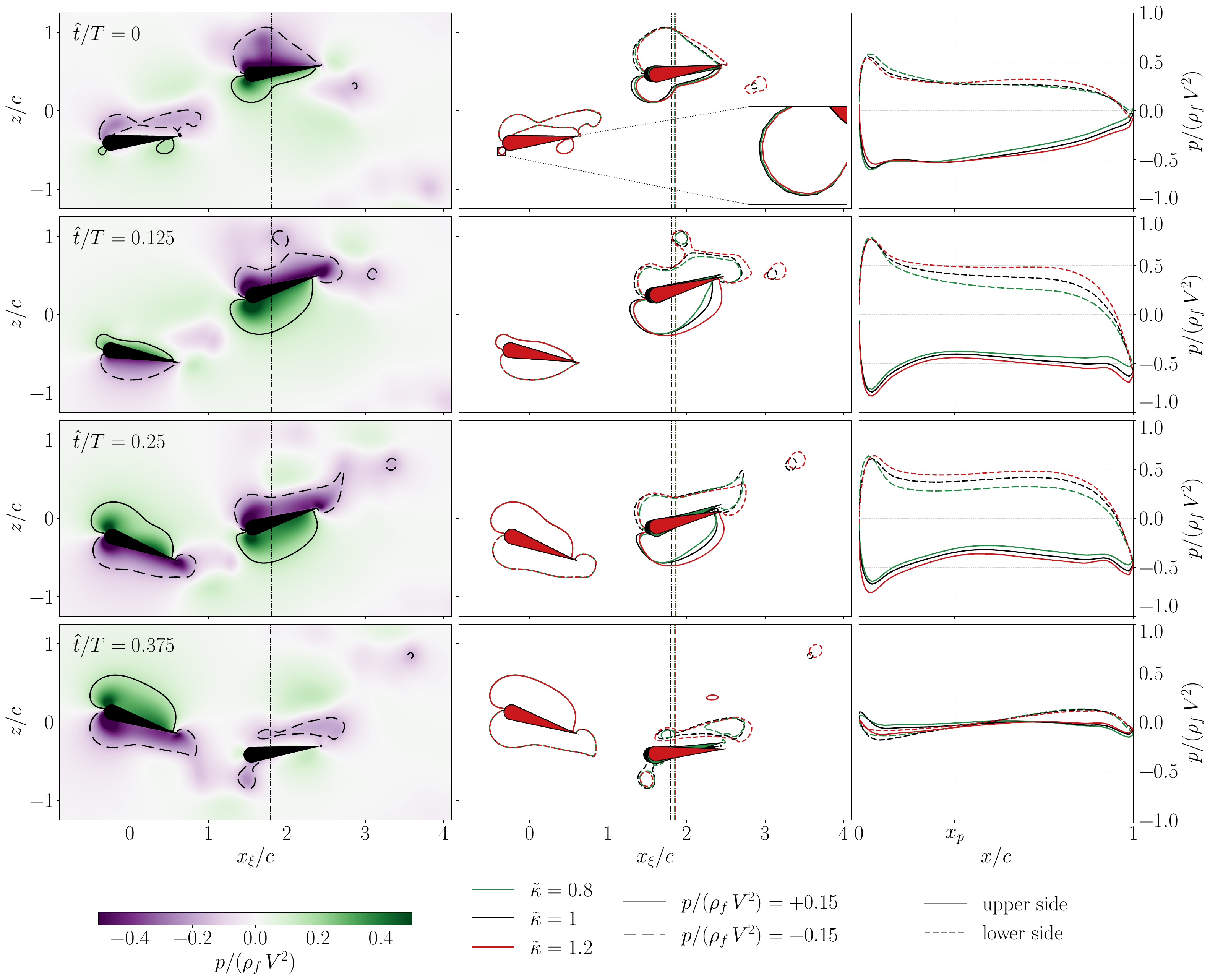}
  \caption{Pressure field around the leader and the follower for the phase
  difference $\gls{dphi} = 135^\circ$ at $\gls{tshift}/\gls{T} = 0.0$, $0.125$,
  $0.25$ and $0.375$ (top to bottom).
  (Left)~pressure field $\gls{p}/(\gls{rhof} \gls{V;sq})$ in the midplane for
  $\gls{knr} = 1.0$, with contours at $\gls{p}/(\gls{rhof} \gls{V;sq}) = \pm 0.15$;
  the dash-dotted line marks the follower's centre of mass.
  (Middle)~contours at $\gls{p}/(\gls{rhof} \gls{V;sq}) = +0.15$ (solid) and
  $-0.15$ (dashed) for $\gls{knr} = 0.8$ (green), $1.0$ (black) and $1.2$ (red).
  The cases are shifted in the streamwise direction by a constant so that
  the mean centres of mass of the leaders coincide; $x_\xi$ is measured from
  the mean centre of mass of the leader for $\gls{knr} = 1.0$, and the
  dash-dotted lines mark the follower's centre of mass in the corresponding
  colour.
  The inset at $\gls{tshift}/\gls{T} = 0$ magnifies the small closed contour
  at the leading edge of the leader: even at this magnification the contours
  of the three cases nearly coincide, showing that the flow around the leader
  is the same in all three simulations.
  (Right)~pressure on the upper (solid) and lower (dashed) surface of the
  follower along the chord, sampled two grid cells from the surface; 
  $\gls{vxp_x}$ marks the centre.
  The dash patterns of the three cases differ only to keep overlapping lines
  distinguishable.}
  \label{fig:res:tandem_opt:power:pressure}
\end{figure}
% -----------------------------------------------------------------------------

% -----------------------------------------------------------------------------
To relate the pitching motion to the input power, we examine the
pressure fields  in Figure~\ref{fig:res:tandem_opt:power:pressure} for
$\gls{dphi}=135^\circ$.
The same qualitative picture is expected to hold for the other phase shifts,
since \gls{knr} shifts $\gls{PiF}$ by a nearly phase-independent amount,
Figure~\ref{fig:res:tandem_opt:power}.
Figure~\ref{fig:res:tandem_opt:power:pressure} shows four instants of the
follower's downstroke, from $\gls{tshift}/\gls{T} = 0$ (top) to $0.375$
(bottom).
The left column shows the pressure $\gls{p}/(\gls{rhof} \gls{V;sq})$ in the 
midplane (\gls{vx_x}--\gls{vx_z}) for the baseline ($\gls{knr} = 1.0$).
The middle column overlays the contours
$\gls{p}/(\gls{rhof} \gls{V;sq}) = \pm 0.15$ of the three stiffness ratios:
the softest follower ($\gls{knr} = 0.8$, green), the baseline (black)
and the stiffest follower ($\gls{knr} = 1.2$, red).
The right column shows the pressure on the upper and lower surface of the
follower along the chord.
The vertical dash-dotted lines mark the follower's centre of mass in the
corresponding colour, and their almost identical position in the three cases
reflects the small influence of \gls{knr} on \gls{D}
(see Figure~\ref{fig:res:tandem_opt:distance}).
% -----------------------------------------------------------------------------

% -----------------------------------------------------------------------------
Around the leader, the contours of the three cases coincide at every instant,
which shows that \gls{knr} does not visibly affect the flow around the
leader.
Around the follower, the contours differ, most clearly at
$\gls{tshift}/\gls{T} = 0.125$ and $0.25$.
At these two instants, the surface pressure magnitude on both sides of the
follower is lowest for $\gls{knr} = 0.8$, and highest for $\gls{knr} = 1.2$.
Hence, the pressure difference across the follower, between its upper and
lower surfaces, decreases as the follower becomes softer.
At $\gls{tshift}/\gls{T} = 0.375$, close to the end of the downstroke of the
follower, the surface pressure is small for all three followers and the
curves collapse.
A smaller pressure difference reduces the vertical hydrodynamic load against
which the actuator works, and hence the required input power.
% -----------------------------------------------------------------------------

% =============================================================================
%   EFFECTIVE ANGLE OF ATTACK
% =============================================================================
\begin{figure}[t!]
  \centering
  \includegraphics[width=0.9\textwidth]{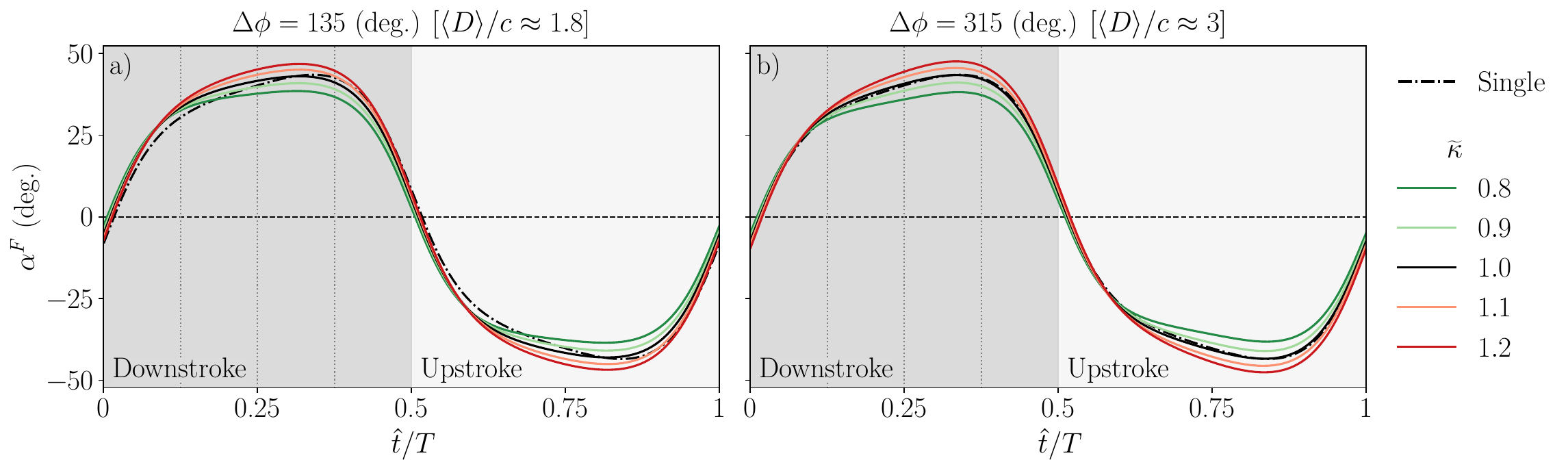}
  \caption{Effective angle of attack of the follower, \gls{aeF},
     as a function of time, for $\gls{knr}=\{0.8,0.9,1,1.1,1.2\}$.
    The dotted vertical lines mark $\gls{tshift}/\gls{T} = 0.125 = 0.125$, 
    $0.25$ and $0.375$, the time instants shown in 
    Figure~\ref{fig:res:tandem_opt:power:pressure}.
    (a) $\gls{dphi}=135^\circ$. (b) $\gls{dphi}=315^\circ$.
  }
  \label{fig:res:tandem_opt:power_phase_aoa}
\end{figure}

% -----------------------------------------------------------------------------
\def\myfig{\ref{fig:res:tandem_opt:power_phase_aoa}}
This is clearly visible by analysing the effective angle of attack,
\begin{equation}
\gls{aeF} = \gls{thF} -
\arctan\left(\frac{\gls{vup_z}}{\gls{Up}}\right)
= \gls{thF} +
\arctan\left(\frac{\sin(2\,\pi\,\gls{f}\,\gls{tshift})\gls{V}}{\gls{Up}}\right),
\end{equation}
shown in Figure~\myfig{}a,b).
With this definition, the effective angle of attack of the follower
accounts for the heaving velocity and the mean propulsive speed,
but not for the velocity induced by the wake of the leader.
The softer the follower, the lower \gls{aeF}, with peak values of
approximately $38^\circ$ for $\gls{knr}=0.8$ and $47^\circ$ for
$\gls{knr}=1.2$.
Since the heaving velocity is fixed and \gls{Up} is nearly
the same for all stiffness ratios, the lower \gls{aeF}
follows mainly from the larger pitching amplitude.
The delay of the pitch peak also contributes to this reduction. For the
softer follower, the minimum of \gls{thF} moves
closer to $\gls{tshift}/\gls{T} = 0.25$ (see
Figure~\ref{fig:res:tandem_opt:power_phase}e,f),
where both the heave-induced angle and the heaving velocity are
largest.
The pitch thus relieves the angle of attack where the heaving motion
is fastest, at the expense of a slightly larger \gls{aeF}  early in
the stroke.

These results identify the mechanism through which the stiffness of the
follower acts.
A softer follower pitches further and slightly later, which lowers its
effective angle of attack and the pressure difference across it.
It therefore requires less input power to sustain the same heaving motion.
In the tandem with equal stiffness, the wake of the leader reduces the
pitching amplitude of the follower and thereby raises its input power.
A softer follower compensates for this reduction and, in the power-saving
cases, pitches with a larger amplitude than the isolated flapper.
% -----------------------------------------------------------------------------

% =============================================================================
% =============================================================================
%   CONCLUSION
% =============================================================================
% =============================================================================

\section{Conclusions} \label{sec:conclusions}
% -----------------------------------------------------------------------------
In this work, we have investigated the self-propulsion and hydrodynamic
interaction of two three-dimensional flappers arranged in an in-line tandem
configuration by means of numerical simulations.
Each flapper self-propels from a prescribed sinusoidal heaving motion imposed
at its centre of mass and it is modelled as a rigid body that pitches freely
about a spanwise axis, restrained by a torsional spring.
The pitching motion is therefore determined by the interaction between the
flapper, the spring, and the surrounding flow.
% -----------------------------------------------------------------------------

% -----------------------------------------------------------------------------
We first set both flappers to the stiffness \gls{knOPT}, which maximises
the propulsive speed of an isolated flapper.
The pair then self-arranges into a stable equilibrium for all phase shifts
($0^\circ \le \gls{dphi} \le 315^\circ$).
The flappers travel at a common propulsive speed, and their equilibrium
distance grows approximately linearly with $\gls{dphi}$.
These distances closely reproduce those of the chordwise-flexible
flappers of \citet{arranz2022} for $90^\circ\le\gls{dphi}\le 225^\circ$,
with moderate differences at the smallest and largest phase shifts.
The leader is essentially unaffected by the follower and requires roughly the
same power as in isolation, whereas the follower requires more power in every
case, with a penalty of up to $22\%$ at $\gls{dphi} = 0$ that decreases
monotonically as the phase difference grows.
Although the tandem configuration propels itself marginally faster than
an isolated flapper (by approximately $3.65\%$), the additional power used
by the follower outweighs this gain, so that the tandem configuration
generally yields no net efficiency benefit over a flapper in isolation.
This contrasts with the chordwise-flexible model of \citet{arranz2022},
which yields a net efficiency benefit in all cases except zero phase
difference.
The power penalty of the follower is associated with a reduction of its
pitching amplitude by the wake of the leader, without a change in pitch timing.
% -----------------------------------------------------------------------------

% -----------------------------------------------------------------------------
Keeping the leader at this stiffness, we have then varied the stiffness of the
follower through the ratio $\gls{knr} = \gls{knF}/\gls{knL}$ in the range
$0.8\le\gls{knr}\le 1.2$.
The formation remains stable for all stiffness ratios considered, and the
equilibrium distance and propulsive speed remain governed primarily by
\gls{dphi}.
Varying $\gls{knr}$ shifts the follower's performance by an almost
phase-independent amount.
For a softer follower ($\gls{knr}<1$), the input power decreases by
up to $8\%$ relative to the untuned case.
For the softest stiffness ratios, this reduction is sufficient for the
follower to require less power than the isolated flapper, from
$\gls{dphi}=135^\circ$ for $\gls{knr}=0.8$ and only at the largest phase
shifts for $\gls{knr}=0.9$.
Stiffer followers, conversely, require up to about 6\% more power and reduce
the efficiency of the tandem.
For $\gls{knr}=0.8$, the tandem is more efficient than an isolated flapper
at every phase shift, with gains of about 2--3\% for $\gls{dphi}\ge 90^\circ$.
Thus, tuning the stiffness of the follower turns a tandem without efficiency
benefit into one that outperforms the isolated flapper across the full phase
range, without active control of the pitching motion.
% -----------------------------------------------------------------------------

% -----------------------------------------------------------------------------
An analysis of the follower over one cycle links this improvement to its
passive pitching response.
A softer follower pitches with a larger amplitude, which compensates for
the reduction imposed by the wake of the leader.
In the cases that require less power than the isolated flapper, this
amplitude even exceeds that of the isolated flapper.
Together with a slight delay of the pitch, the larger amplitude lowers the
effective angle of attack of the follower during the peaks of input power.
The resulting smaller pressure difference reduces the normal force on the
follower, and hence the vertical load against which the heaving motion is
sustained.
Because the softer follower is more tilted, the streamwise part of this force
is maintained, consistent with the nearly unchanged propulsive speed.
% -----------------------------------------------------------------------------

% -----------------------------------------------------------------------------
These conclusions apply to a single set of parameters (density ratio, Reynolds
number, aspect ratio and amplitude of motion were kept fixed) with lateral
motion constrained.
Preliminary simulations indicate that the benefit of softening the
follower is bounded, since for $\gls{knr}\lesssim0.7$ its drag exceeds its
thrust and the equilibrium distance grows rapidly.
The present study has considered only an in-line tandem of two flappers.
Extending it to lateral or staggered arrangements and to larger groups would
show how far the simplified structural model carries over to collective
locomotion.
% -----------------------------------------------------------------------------

% =============================================================================
% =============================================================================
%   OUTRO
% =============================================================================
% =============================================================================
\section*{Statements and Declarations}

\textbf{Authors’ Individual Contributions}
Conceptualization: GA, OF, MM, MGV;
Methodology: GA, OF, MM, MGV;
Investigation: MS;
Formal analysis: MS;
Visualization: MS;
Software: MS, MM;
Writing – original draft preparation: MS;
Writing – review and editing: GA, OF, MM, MGV;
Supervision: MM, MGV.

\textbf{Statement on the use of generative AI}
 During the preparation of this work, the authors used Claude (Anthropic) in 
 order to improve the readability and flow of the text as well as to check 
 for consistency of the text. 
 After using this tool, the authors reviewed and edited the content as needed 
 and take full responsibility for the content of the published article.

%\textbf{Acknowledgments}

\bibliographystyle{apalike}   
\bibliography{reference.bib}

\end{document}